\documentclass{article}

\usepackage{PRIMEarxiv}

\usepackage[utf8]{inputenc} % allow utf-8 input
\usepackage[T1]{fontenc}    % use 8-bit T1 fonts
\usepackage{hyperref}       % hyperlinks
\usepackage{url}            % simple URL typesetting
\usepackage{booktabs}       % professional-quality tables
\usepackage{amsfonts}       % blackboard math symbols
\usepackage{nicefrac}       % compact symbols for 1/2, etc.
\usepackage{microtype}      % microtypography
\usepackage{lipsum}
\usepackage{fancyhdr}       % header
\usepackage{graphicx}       % graphics
\graphicspath{{media/}}     % organize your images and other figures under media/ folder

\usepackage{color}
\usepackage{xcolor}
\usepackage{listings}
\usepackage{lineno}

\definecolor{codegreen}{rgb}{0,0.6,0}
\definecolor{codegray}{rgb}{0.5,0.5,0.5}
\definecolor{codepurple}{rgb}{0.58,0,0.82}
\definecolor{backcolour}{rgb}{0.95,0.95,0.92}

\lstdefinestyle{mystyle}{
    backgroundcolor=\color{backcolour},
    commentstyle=\color{codegreen},
    keywordstyle=\color{blue},
    numberstyle=\tiny\color{codegray},
    stringstyle=\color{codepurple},
    basicstyle=\ttfamily\scriptsize,
    numbers=left,
    numbersep=5pt,
    breaklines=true,
    breakatwhitespace=true,
    showstringspaces=false,
    captionpos=b,
    frame=single
}

\title{Probeable Problems for Beginner-level Programming-with-AI Contests
}

\author{
  Mrigank Pawagi\\
  Indian Institute of Science \\
  Bengaluru\\
  \texttt{mrigankp@iisc.ac.in} \\
   \And
  Viraj Kumar \\
  Indian Institute of Science \\
  Bengaluru\\
  \texttt{viraj@iisc.ac.in} \\
}

\newcommand{\pr}{Probeable}
\newcommand{\makelink}[2]{\href{#1}{\textcolor{blue}{[\underline{#2}]}}}

\begin{document}
\maketitle

\begin{abstract}
To broaden participation, competitive programming contests may include beginner-level problems that do not require knowledge of advanced Computer Science concepts (e.g., algorithms and data structures). However, since most participants have easy access to AI code-generation tools, these problems often become trivial to solve. For beginner-friendly programming contests that do not prohibit the use of AI tools, we propose \emph{\pr} Problems: code writing tasks that provide (1)~a problem specification that deliberately omits certain details, and (2)~a mechanism to probe for these details by asking clarifying questions and receiving immediate feedback. To evaluate our proposal, we conducted a 2-hour programming contest for undergraduate Computer Science students from multiple institutions, where each student was an active member of their institution's computing club. The contest comprised of six {\pr} Problems for which a popular code-generation tool (GitHub Copilot) was unable to generate accurate solutions due to the absence of details. Students were permitted to work individually or in groups, and were free to use AI tools. We obtained consent from 26~groups (67~students) to use their submissions for research. We analyze the extent to which the code submitted by these groups identifies missing details and identify ways in which {\pr} Problems can support learning in formal and informal CS educational contexts.
\end{abstract}

% keywords can be removed

\section{Introduction}
Participants in competitive programming contests typically solve a series of code-writing tasks within a stipulated time. The premier competitive programming contest is the International Collegiate Programming Contest (ICPC), whose purpose\footnote{\url{https://icpc.global/worldfinals/fact-sheet/ICPC-Fact-Sheet.pdf}} is ``to advance prospects for the next generation by bringing students together, working collaboratively to solve \emph{algorithmically challenging problems}, and preparing them to build dependable systems that benefit their neighbors with the support of universities, industry, and community leaders, globally'' (emphasis added). To welcome beginners to the exciting world of competitive programming, many contests, including those organized by popular websites such as LeetCode, AtCoder, and CodeForces, have traditionally offered beginner-level code-writing problems that require no knowledge of advanced Computer Science concepts in algorithms and data structures. Today, most participants have easy access to AI code-generation tools that can often solve beginner-level code-writing problems~\cite{10.1145/3511861.3511863, 10.1145/3576123.3576134, denny_sigcse23, 10.1145/3568813.3600142, venkatesh2023suppressed}. While the use of such tools may be banned by some contest organizers, others may permit them because they are unwilling or unable to prohibit their use.
Thus, our study is motivated by the following question:
\begin{quote}
Is it possible to design beginner-level competitive programming problems that \emph{cannot} be solved successfully by AI code-generation tools alone?
\end{quote}
For instance, since the large language models (LLMs) underlying modern code-generation tools are expensive to retrain and are known to be sensitive to subtle changes in the format of their prompts~\cite{sclar2023quantifying}, it may be feasible to create novel (or seemingly-novel) variants of beginner-level problems to foil such tools. In a recent study, Huang et al.~\cite{huang2023competitionlevel} showed that GPT-4 excels at ``easy'' code-generation problems from CodeForces contests conducted prior to September 2021 (the cut-off for GPT-4's training\footnote{\url{https://platform.openai.com/docs/models/gpt-4-and-gpt-4-turbo}}), but this performance falls sharply thereafter. The authors define ``easy'' problems as those between 800 to 1,100 on the CodeForces difficulty ratings for problems. However, Jain et al.~\cite{jain2024livecodebench} argue that beginner-level (Divisions 3 and 4) CodeForces problems are more challenging than ``easy'' problems from LeetCode and AtCoder. Hence, they define ``easy'' CodeForces problems as those with the lowest possible rating (800), and ``hard'' problems as those rated over 1,000. Based on this scale, Jain et al. maintain a live leaderboard showing code-generation performance of several LLMs on newly released contest problems\footnote{\url{https://livecodebench.github.io/leaderboard.html}}. At the time of writing, the best-performing model (GPT4-Turbo) successfully solves over 80\% of ``easy'' problems from the most recent contests.

\begin{figure*}[t]
  \centering \includegraphics[width=\textwidth]{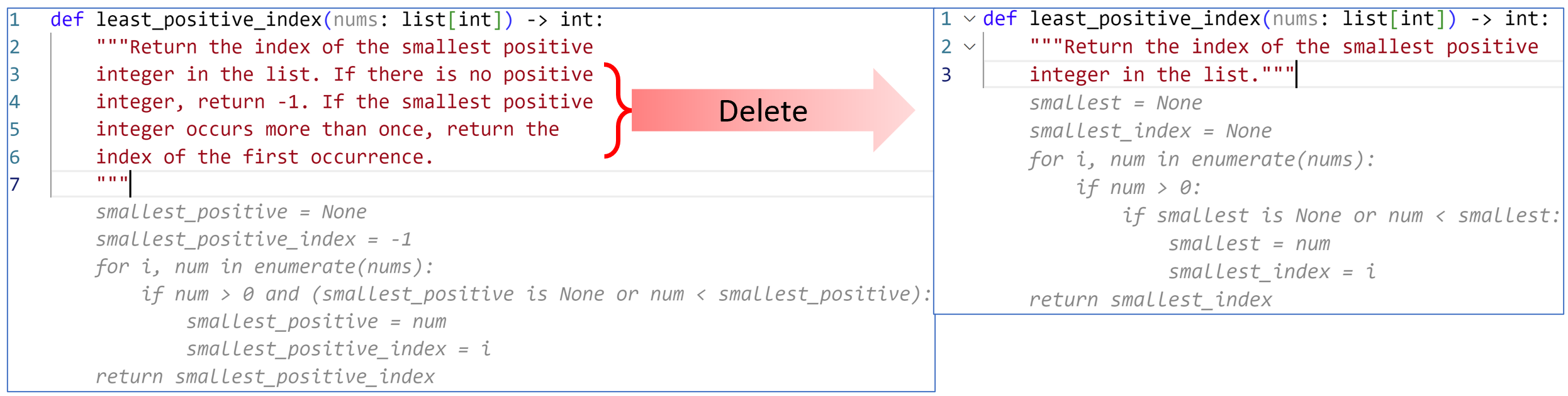}
  \caption{(Left, full specification) A screenshot showing a correct solution suggested by GitHub Copilot (grey font, after Line 7). (Right, after trimming away details) GitHub Copilot suggests a functionally \emph{inequivalent} solution (after Line 3), since it is forced to make ``reasonable'' decisions about the omitted details.}
  \label{fig:method}
\end{figure*}

To avoid ambiguity, the task specifications for competitive programming problems are fully detailed. In contrast, real-world software specifications can be ambiguous in a variety of ways~\cite{berry2004, dupre1995bugs, kovitz1998practical}. To successfully thwart AI code-generation tools on beginner-level code-writing tasks, we draw inspiration from two sources. The first source is a specific type of ambiguity that Schneider called for in 1978~\cite{schneider1978}. He argued that CS1 students ``should be able to recognize and resolve uncertainties in simple problem statements''. Further, as one way to develop this ability, he suggested that ``some programming assignments should \emph{intentionally be left incomplete}, requiring the student to \emph{consider the alternatives} and to \emph{make a reasonable decision} on the omitted details'' (emphasis added). Our second source of inspiration is the eagerness that current AI tools display in generating code even when the specification is incomplete. Figure~\ref{fig:method} shows that an AI code-generation tool (GitHub Copilot) can easily generate a correct solution when presented with the full specification (left). This is consistent with prior research on the effectiveness of tools like GitHub Copilot when they are provided full specifications for CS1 code-writing problems~\cite{denny_sigcse23, wermelinger2023sigcse}. However, when key details from this specification are omitted, the tool nevertheless attempts to generate code. In doing so, it makes \emph{silent} decisions (which may or may not be reasonable in Schneider's sense) to resolve the ambiguities created by omitting details.

\subsection{{\pr} Problems}\label{sec:pr_intro}
A {\pr} Problem has two components: (1)~a code-writing task to create a function given an \emph{incomplete} specification, and (2)~an oracle that reveals the function's desired behavior on any input(s) chosen by the contestant. The contestant must recognize on their own that the specification allows more than one \emph{possible} result on a certain input before asking the oracle for the desired result on that input. While the oracle denies contestants an opportunity to make their own decisions regarding the details omitted by the specification, we believe that this is necessary for two reasons: it imposes no additional burden on contest organizers, and it keeps the problem difficulty level the same for all contestants.

\begin{figure}
  \includegraphics[width=\textwidth]{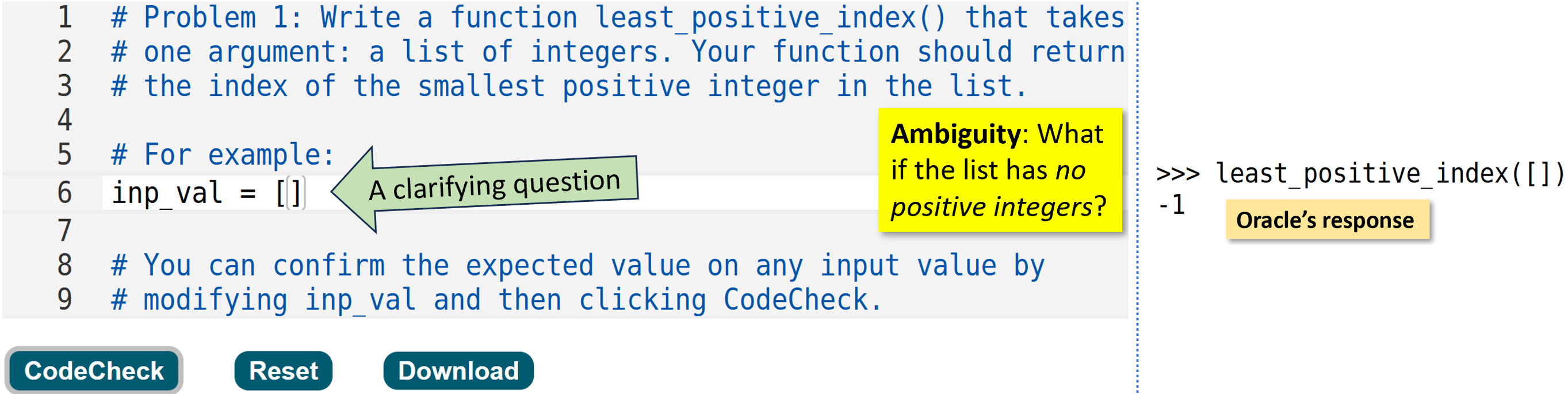}
  \caption{(Left) A \emph{\pr} Problem with omitted specification details, and a contestant's \emph{clarifying question} attempting to resolve one such detail. (Right) The automated feedback generated by an `oracle' specifies the desired output on this input.}
  \label{fig:teaser}
\end{figure}

Figure~\ref{fig:teaser} shows an example of a {\pr} Problem, whose task specification deliberately omits two details. First, it does not specify what the function should return when the input list contains no positive integers (e.g., when it is empty). Second, it does not specify which index the function should return if the smallest positive integer occurs more than once in the list. When presented with such a specification, what might a contestant do? We consider three among several possibilities in the following vignettes.

\paragraph{Fictitious Contestant 1}
This contestant fails to recognize that the specification lacks detail. Next, as shown in Figure~\ref{fig:method} (right), they rewrite the given information as a function signature and docstring and they generate a solution using an AI code-generation tool. Finally, they submit this solution after satisfying themselves that it corresponds to their (incorrect) understanding of the task.

\paragraph{Fictitious Contestant 2}
This contestant mimics Contestant~1 up to the point where they have generated the code in Figure~\ref{fig:method} (right). Then, while examining the generated code, they realize that it returns \texttt{None} when the input is an empty list. Further, they review the specification and confirm that it fails to specify the return value on this input. Next, they ask the oracle for the desired return value on input \texttt{[]}, as shown in Figure~\ref{fig:teaser}. They replace the AI-generated statement \texttt{smallest\_index = None} with \texttt{smallest\_index = -1} and submit the code, believing that they have solved the problem correctly. (They have not.)

\paragraph{Fictitious Contestant 3}
This contestant immediately recognizes that the function's desired return value is unspecified for the common ``corner case'' input \texttt{[]}. They query the oracle as above and then mimic Contestant~1 by writing the function's signature and docstring. However, based on the oracle's output, they add an additional sentence to the docstring: If the list is empty, return -1. Next, they use an AI tool to generate code that is identical to the code in Figure~\ref{fig:method} (right) except that \verb|smallest_index| is initialized to \texttt{-1} instead of \texttt{None}. While reviewing the code, they realize that the specification was incomplete not only for the specific list \texttt{[]}, but for the whole category of lists that contain no positive integers. Thus, they query the oracle again, this time on the input \texttt{[0]}. The oracle reports the desired return value as \texttt{-2}. Somewhat mystified, the contestant queries the oracle again with inputs \texttt{[-1]} (return value: \texttt{-2}) and \texttt{[0, -1]} (return value: \texttt{-3}). At this point, the contestant inductively guesses a pattern: the return value is \texttt{-(len(nums) + 1)}. They use this expression in place of the \texttt{-1} in the AI-generated \texttt{return} statement. They now submit the code, failing to recognize that the specification was incomplete in another respect that causes their code to fail on inputs such as \texttt{[1, 1]}.

\subsection{Designing {\pr} Problems}\label{sec:design_pr}
For our study, we have attempted to design a set of {\pr} Problems that omit details of diverse types. We used one of these as a Practice Problem to introduce contestants to this style of problem, as described in Section~\ref{sec:contest}. The remaining five {\pr} Problems were used for the contest. Prior to the contest, we ensured that none of these popular AI code-generation tools was able to generate correct solutions for these problems, when prompted as shown in Figure~\ref{fig:method} (right): GitHub Copilot, ChatGPT-3.5, and Codeium.

In our initial experiments with designing {\pr} Problems, we found that some students started querying the oracle for the expected behaviour on unusual inputs (e.g., strings or lists containing floating-point numbers for a function whose argument was a list of integers). To limit the scope of such exploration, we have designed our oracles to give an appropriate error message on such inputs.
%\footnote{We plan to release a template for designing such problems in the final version of our paper.}

Since we are unaware of prior work that categorizes various types of omitted code-writing specification details, we draw from our own experience in failing to design fully-specified code-writing tasks (e.g., for traditional code-writing assignments). Our six {\pr} Problems illustrate four types of omissions. For each problem, we began our design with a specification for a code-writing task that was fully detailed (to the best of our knowledge), and then judiciously omitted\footnote{See Section~\ref{sec:testing} for prior work that influences our selection of which details to omit.} some details of the following types:

\subsubsection{Definitions}
\begin{figure}
    \centering
    \includegraphics[width=0.75\textwidth]{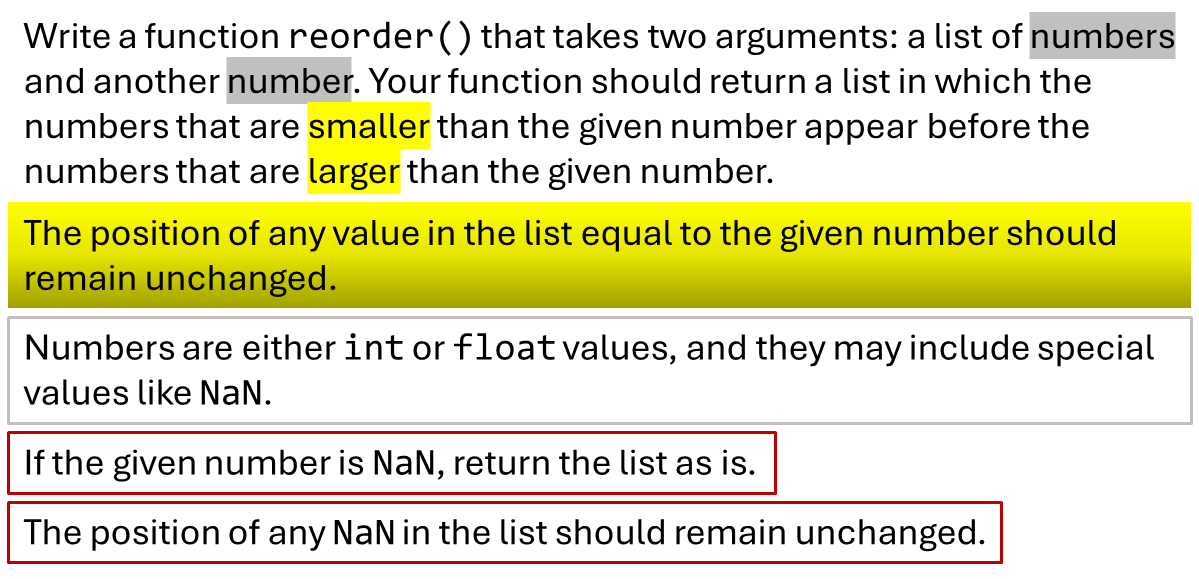}
    \caption{Problem 5. The task specification draws attention to values that are ``smaller'' and ``larger'' than the given number (highlighted in plain yellow) but deliberately neglects details on handling values equal to the given number (highlighted in chiaroscuro yellow). The specification also deliberately omits a definition of the grey-highlighted terms (shown in the grey box) and the return value when the arguments involve \texttt{NaN} (shown in the red boxes). CodeCheck: \makelink{https://codecheck.io/files/23052009254fsf3krf2kbaao8g69hezp1xw}{oracle}, \makelink{https://codecheck.io/files/23122108311syqwhqt1v4gf6t3wzx6lbmqs}{verifier}}
    \label{fig:p5}
\end{figure}

In each of four problems, we entirely omit the detailed definition of one key term from the specification. For the Practice Problem, this term is ``palindrome''. (For clarity, this term is highlighted in grey in Figure~\ref{fig:prac} and the omitted definition is shown in a grey box.) We expect CS students (and LLMs) to recall a familiar definition of this term: a string that is precisely equal to its reverse (e.g., \verb|'racecar'|). However, the omitted details expand this definition to include strings such as \verb|'race car'| and \verb|'Racecar'|. In contrast, we do not expect CS students or LLMs to associate the grey-highlighted terms in Problem~2 (Figure~\ref{fig:p2}) and Problem~4 (Figure~\ref{fig:p4}) with any specific definition. In Problem~5 (Figure~\ref{fig:p5}), the grey-highlighted term ``number'' frequently refers to either integer or floating-point values, and we omit the fact that it refers to both types here.

\subsubsection{Behavior on certain inputs}
Ideally, the specification should describe the function's expected behavior on each possible type of input. However, we sometimes fail to describe the expected behavior on certain types of inputs. For Problem~4, one such omission corresponds to a specific input (the empty list). We show such an omission within a red box with dashed lines (Figure~\ref{fig:p4}). A student who recognizes such an omission can easily ask the oracle for the expected behavior on that specific ``corner case'' input.

In contrast, if the missing details correspond to a whole class of inputs, these omissions are usually shown within a red box with solid lines. (An exception is described in the next paragraph.) For example, the given Problem~1 specification (Figure~\ref{fig:p1}) omits details not only for the empty list but for any input list that has no positive integers (e.g., \texttt{[0, -1]}). Since the oracle only reveals the expected behavior on \emph{specific} inputs, in such cases contestants must query the oracle often enough to \emph{inductively guess} the expected behavior on all inputs for which details are omitted.

In our experience, we have often observed a particular sub-type of this type of omission: when the given specification explicitly describes the expected behavior for a certain category of inputs but neglects to do so for a closely related category. This reminds us of the artistic term chiaroscuro\footnote{\url{https://en.wikipedia.org/wiki/Chiaroscuro}}, and we use an alternate form of highlighting for such omissions. As an example, the given specification for Problem~4 (Figure~\ref{fig:p4}) draws attention to \emph{positive} integers in the list (highlighted in plain yellow), but fails to clarify the expected behavior when the list contains \emph{non-positive} values (highlighted in chiaroscuro yellow).

\subsubsection{Return type(s)}
\begin{figure}
    \centering
    \includegraphics[width=0.75\textwidth]{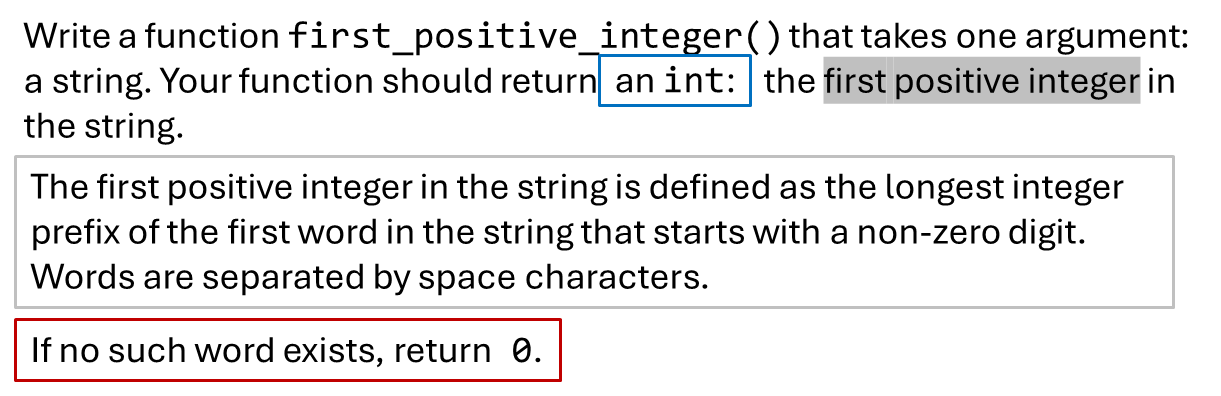}
    \caption{Problem 2. The task specification deliberately omits the return type (shown in the blue box) and a definition for the grey-highlighted term ``first positive integer''. This definition is non-trivial, and is shown in the grey box. The value to be returned when the string has no such positive integer (shown in the red box) is also deliberately omitted. CodeCheck: \makelink{https://codecheck.io/files/2306111051595nfjvjxiu7a73md5cn4saj9}{oracle}, \makelink{https://codecheck.io/files/23122108113rh4yb9fq33jmta2vlozb9nxe}{verifier}}
    \label{fig:p2}
\end{figure}

In two problems, we omit details that would clarify the function's exact return type(s). In Problem~2, the desired return type could conceivably have been \texttt{str} instead of the omitted type \texttt{int} (shown within a blue box in Figure~\ref{fig:p2}). Similarly, the two values \texttt{buy} and \texttt{sell} could have been returned as a \texttt{list} instead of a \texttt{tuple}, as shown in Figure~\ref{fig:p4}. This function has two return types: on the empty list, it returns \texttt{None} (of type \texttt{NoneType}).

\subsubsection{Tie-break mechanism(s)}
\begin{figure}
    \centering
    \includegraphics[width=0.75\textwidth]{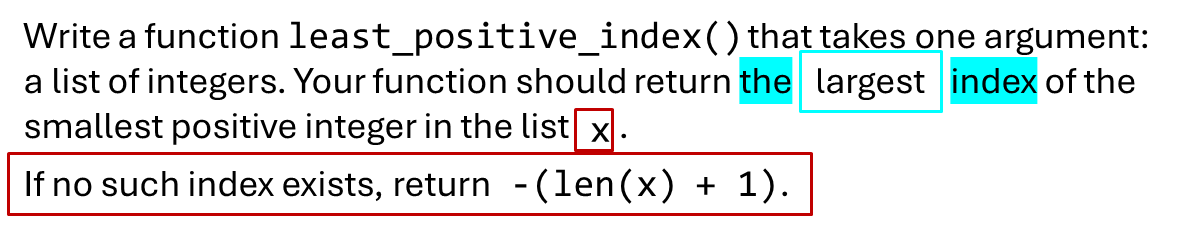}
    \caption{Problem 1. The task specification refers to ``the index'' (highlighted in cyan) without clarifying how to break ties, if any. The tie-break mechanism ``largest'' (shown in the cyan box) is deliberately omitted. The value to be returned when the list has no positive integers (shown in the red box) is also deliberately omitted. CodeCheck: \makelink{https://codecheck.io/files/2306111033cnnmzafkxveg0ap6i7blj01f0}{oracle}, \makelink{https://codecheck.io/files/2312210759einuagsjv9d9v447safopvrs7}{verifier}}
    \label{fig:p1}
\end{figure}

\begin{figure}
    \centering
    \includegraphics[width=0.75\textwidth]{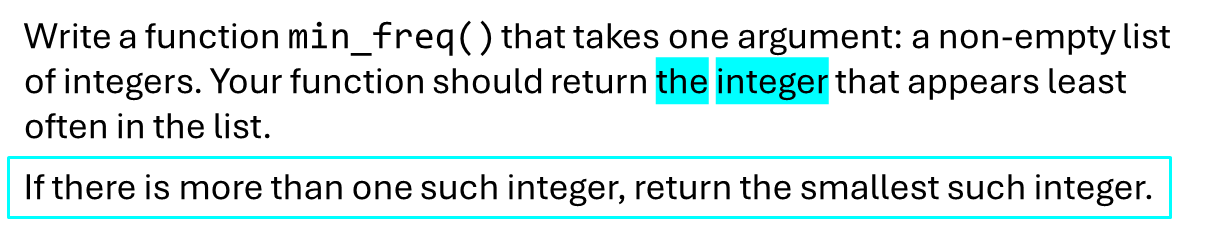}
    \caption{Problem 3. The task specification refers to ``the integer'' (highlighted in cyan) without clarifying how to break ties, if any. The tie-break mechanism (shown in the cyan box) is deliberately omitted. CodeCheck: \makelink{https://codecheck.io/files/231221074114f9xinsv2s4qoe8ekbcp6cng}{oracle}, \makelink{https://codecheck.io/files/2312210814ayo7lh088ko8jv0whrgdbeei6}{verifier}}
    \label{fig:p3}
\end{figure}

\begin{figure}
    \centering
    \includegraphics[width=0.75\textwidth]{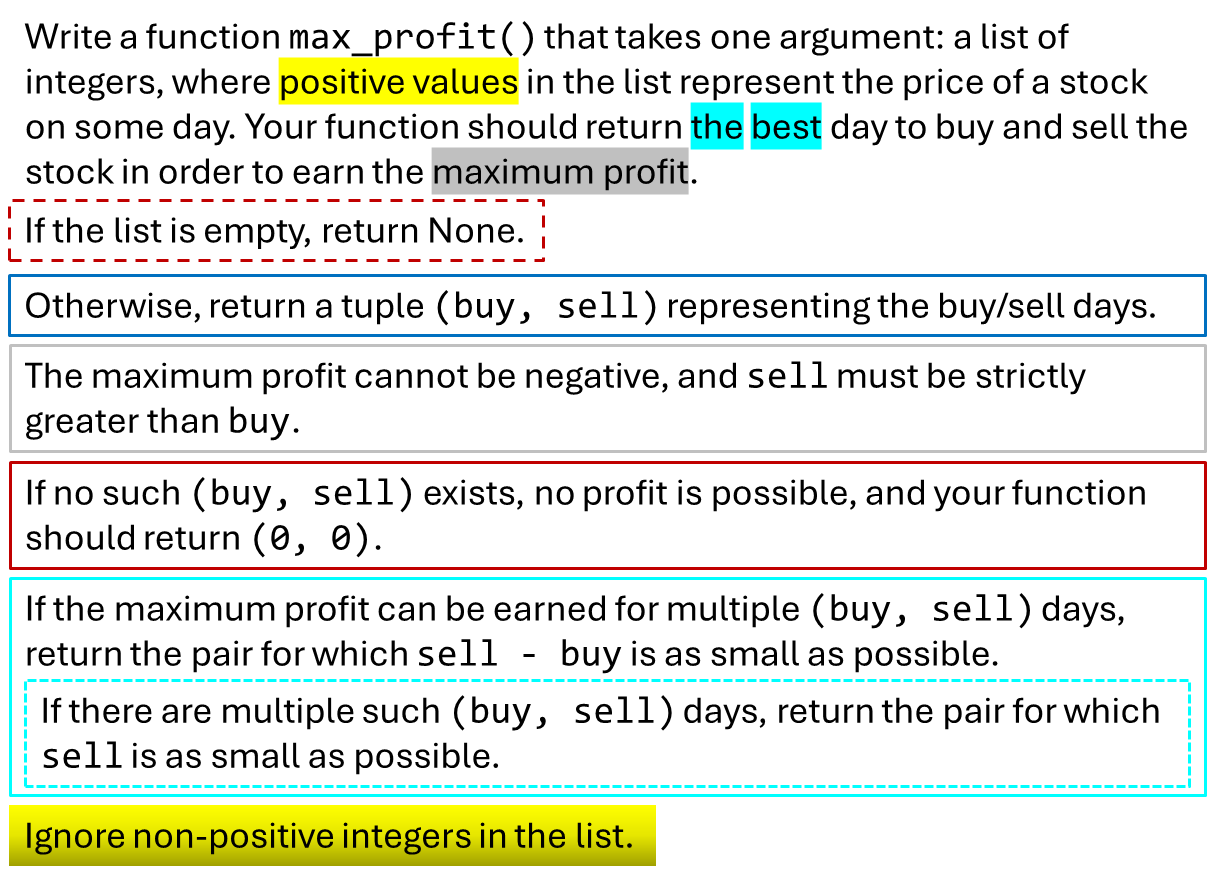}
    \caption{Problem 4. The task specification draws attention to ``positive values'' in the list (highlighted in plain yellow) but deliberately neglects details on handling non-positive values (highlighted in chiaroscuro yellow). The \emph{specific} value to be returned when the list is empty (shown in the red box with dashes) is deliberately omitted. The specification also deliberately omits the return type (shown in the blue box), the definition of the grey-highlighted term ``maximum profit'' (shown in the grey box), the return value when only non-negative profit can be earned (shown in the red box), and the tie-break mechanism for the cyan-highlighted term ``the best'' (a complex, nested mechanism shown in the cyan box). CodeCheck: \makelink{https://codecheck.io/files/23052001283if0mgoiorxweda5phl1u4769}{oracle}, \makelink{https://codecheck.io/files/23122108196i0eervhvcevuat25j6nnyzgq}{verifier}}
    \label{fig:p4}
\end{figure}

In each of three problems, the definite article \emph{the} is used by the given specification to \emph{uniquely} describe the expected behavior for inputs with specific properties. However, for inputs where these properties do not hold, more than one behavior is possible and the specification neglects to clarify how to break ties in such cases. For example, in Problem~3 (Figure~\ref{fig:p3}), the cyan-highlighted expression ``the integer'' is unambiguous only for lists where a unique integer appears least often. The specification omits details (shown in the cyan box) that describe a tie-break mechanism when more than one integer appears least often in the list. While the tie-break mechanism can sometimes be described in just one word (e.g., for Problem~1 in Figure~\ref{fig:p1}), it may be extremely complex (e.g., for Problem~4 in Figure~\ref{fig:p4}). In such cases, it may be challenging for contestants to probe the oracle with inputs that correspond to various tie-break scenarios.

\begin{figure}
    \centering
    \includegraphics[width=0.75\textwidth]{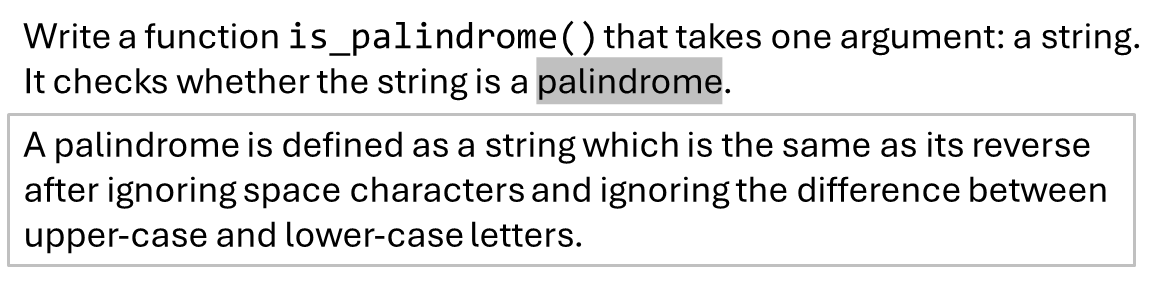}
    \caption{Practice Problem. The given task specification deliberately omits the definition of ``palindrome'' (highlighted in grey). The omitted definition is non-standard (shown in the grey box). CodeCheck: \makelink{https://codecheck.io/files/23122107005thqqy15kp8cl3u9pdekwrs01}{oracle}, \makelink{https://codecheck.io/files/23122107523v0rdcv2x6bpktj02n5ke7ovw}{verifier}.}
    \label{fig:prac}
\end{figure}

\subsection{Research Questions}
Since our questions are designed for a competitive programming contest, they are meant to be challenging. However, as we describe in Section~\ref{sec:methodology}, submissions were evaluated based on which omitted details contestants were successfully able to account for in their code. We expect certain types of omissions (e.g., return types or the expected behavior on \emph{specific} inputs) to be easier to probe for than others (e.g., definitions). Even for the same type of omission, the difficulty of understanding the expected behavior via probing is likely to depend on the nature of the problem. For example., the tie-break condition for Problem~1 (Figure~\ref{fig:p1}) is likely to be far easier to deduce than the tie-break condition for Problem~4 (Figure~\ref{fig:p4}). Hence, we analyze the performance at the level of each (problem, omission) pair.

For each {\pr} Problem, contestants are obliged to probe the oracle on a series of carefully crafted inputs before they can inductively guess the expected behavior, as illustrated in the vignette for Fictitious Contestant~3 (Section~\ref{sec:pr_intro}). As illustrated in that vignette, contestants may succeed in inductively guessing the missing details for one omission, but fail to recognize further omitted details. To provide some assistance, contestants were given access to verifiers for each {\pr} Problem after submitting their first attempts. (Further details on these verifiers is deferred to Section~\ref{sec:methodology}). Contestants could use these verifiers as often as they wished to check the correctness of subsequent versions of their code, before submitting their second (and final) attempts.

Our research questions are as follows:

\begin{description}
\item[RQ1] How successful were contestants at determining the missing details for each (problem, omission) pair?

\item[RQ2] Were contestants able to leverage the verifiers to identify additional omissions?
\end{description}

\section{Related Work}
We position our work in relation to three major strands of prior research. First, since many educators are concerned that beginners will develop an over-reliance on AI code-generators~\cite{10.1145/3545945.3569759} and may lose their coding skills~\cite{denny2023computing} (skills that competitive programming has traditionally sought to bolster), we consider alternatives to traditional textual specifications that have been proposed for code-writing tasks (Section~\ref{sec:alt_spec}). Next, we contrast competitive programming contests with {\pr} Problems against other types of competitive events that also expect contestants to consider the behavior of code on a variety of inputs (Section~\ref{sec:alternatives}). Finally, in Section~\ref{sec:testing} we compare our approach of allowing students to test their understanding of the specification using oracles with other ways of testing such understanding that have been proposed. We also review prior work on how students develop tests, and how educators can help students develop more effective tests.

\subsection{Alternative Specifications}\label{sec:alt_spec}
As noted previously, a number of
AI code-generation tools are easily available for students (e.g., ChatGPT, GitHub Copilot). These tools can often translate natural-language specifications of beginner-level code-writing tasks into accurate solutions~\cite{10.1145/3511861.3511863, 10.1145/3576123.3576134, denny_sigcse23, 10.1145/3568813.3600142, venkatesh2023suppressed}. As tasks grow in complexity, these tools can continue to generate accurate solutions, particularly if students decompose these tasks into simpler sub-tasks~\cite{denny_sigcse23}. Significantly more advanced code-generation tools are being developed. Since these are less readily available to students at present, we defer a discussion of these tools to Section~\ref{sec:advanced_ai}.

To foil the usage of such AI-generation tools, researchers have recently experimented with replacing textual specifications with visual specifications. Rojit et al. originally proposed such specifications for web-application programming to aid specification comprehension in contexts where ``learners or instructors lack fluency in the language of communication''~\cite{rojit2016visual}. More recently, visual specifications have been used to design Prompt Problems~\cite{denny2024promptproblems}, where an illustration shows how specific input values should be converted into outputs. To solve such a problem, students submit a general prompt that is inductively deduced from these examples~\cite{prather2024interactions}. The quality of this submission is assessed by using a code-generator to translate the generalized prompt into code, which is then compared against the instructor's reference solution.

There are two reasons why specifications that represent input-output examples visually are unlikely to foil AI code-generation tools in the context of beginner-level competitive programming. First, Ventakesh et al.~\cite{venkatesh2023suppressed} show that many such problems can be solved using just meaningful names of functions and arguments, and a small number of input-output examples. The latter can be easily transcribed from visual specifications. Second, as visual transformers and other multi-modal foundation models continue to improve~\cite{xu2024survey}, students may gain access to AI-tools that can directly convert visual specifications into code.

Further, unlike visual specifications where the examples are pre-defined, the oracle for a {\pr} Problem allows students to clarify the expected behavior on any input. In this respect, {\pr} Problems are similar to Executable Examples~\cite{icer2019}. This connection is explored further in Section~\ref{sec:testing}.

\subsection{Alternatives to Competitive Programming}\label{sec:alternatives}
While competitive programming typically focuses on writing code given detailed task specifications, prior work explored an alternative contest form called SCORE that touches ``all facets of software development, from requirements analysis and elicitation to implementation and verification''~\cite{mandrioli2010score}. Similar to {\pr} Problems, tasks in a SCORE contest are poorly specified. To obtain clarity, contestants must frequently interact with a \emph{human} ``stakeholder'' (vs. an automated ``client'' oracle in our case), and this stakeholder plays a key role in ensuring that contestants devote ``considerable attention to user needs and requirements elicitation''~\cite{mandrioli2010score}. Further, a SCORE contest is organized over several weeks and targets contestants with some background in software engineering. Thus, this format is unsuitable for short contests that target beginners.

We further compare competitive programming contests using {\pr} Problems with two other types of competitive events: the well-known Capture the Flag contests (which are used in informal educational settings, competitions, and more recently, in formal education contexts~\cite{10.1145/3328778.3366893}) and Competitive Debugging~\cite{rein2022competitive}. In all these types of contests, participants seek to develop an understanding of a given system by carefully choosing certain inputs and examining the system's behavior before writing (or generating) code. {\pr} Problems seem to require a mindset more akin to CTF, since contestants are trying to find inputs that expose previously unknown or unexpected behaviors, whereas debugging usually begins with at least one known failure-inducing input.

\subsection{Understanding the Specification}\label{sec:testing}
{\pr} Problems explicitly require contestants to reason about the (inadequacies of the) given specification or requirements. This skill may be under-developed, even amongst professionals. For instance, Aniche et al.~\cite{9625808} note that developers reasoning about the adequacy of their test cases tend to focus on their source code, their test code, and its code coverage while often failing to consider the documented specification. Such a biased focus while composing test suites or test plans may reflect practices developed during their formal education, where emphasis is often given to test adequacy criteria such as branch coverage or mutation analysis (e.g.,  \cite{10.1145/2648787}, \cite{10.1145/3287324.3287366}).

In a fascinating study of graduate students tasked with testing a single problem, Doorn et al. observed that the tendency to think from an ``implementation-first'' mindset can appear even when \emph{no implementation is provided}, resulting in a failure to adequately cover the problem space.

To counteract this mindset, Shin and Kazerouni~\cite{shin2024model} encourage a ``requirements-first'' approach to testing. {\pr} Problems align well with this goal by encouraging contestants to ask clarifying questions related to the requirements.

{\pr} Problems that fail to specify the behavior on certain inputs (particularly the chiaroscuro sub-type that draw attention to a subset of the input space) may be challenging for beginners, who often consider test inputs only for the happy path~\cite{10.1145/3430665.3456368, 10.1109/ICSE-SEET52601.2021.00029, 10.1145/2591708.2591757}. Similarly, {\pr} Problems that omit details for certain ``corner case'' inputs may challenge beginners, who often fail to consider such inputs while write test cases~\cite{10.1145/3287324.3287461, 10.1109/ICSE-SEET52601.2021.00029}. 

To help students develop their ability to write effective tests, Cordova et al.~\cite{10.1145/3408877.3432417} and Wrenn et al.~\cite{icer2019} have developed feedback mechanisms that provide students with positive reinforcement when they explore conceptually interesting regions within the input space. In a similar spirit, we have attempted to select conceptually interesting details to omit from our specifications while developing {\pr} Problems for this study.

Our problem specifications contain some natural language that students must comprehend. Wrenn et al. showed that an oracle, Examplar, can improve students' comprehension of the specification~\cite{icer2019, wrenn2021reading}. Further, since instructors may \emph{inadvertently} introduce ambiguities into code writing specifications by neglecting certain details, Examplar may be helpful in identifying such missing details~\cite{prasad2022making_hay}. Our approach differs from this line of research in two ways. First, we consciously and carefully introduce ambiguities by omitting key details (see Section~\ref{sec:design_pr}). Second, Examplar asks a student to specify both an input and the output they expect on that input. Since this oracle's purpose is to assist students in comprehending the specification, it indicates whether or not the output expected by the student matches the desired output for that input. In contrast, the purpose of the oracle in a {\pr} Problem is to resolve the ambiguities that have been deliberately introduced by omitting details. An Examplar-like oracle may force contestants to blindly guess the desired output, which could be frustrating.

\section{Study Context}\label{sec:contest}
As a first evaluation of {\pr} Problems, we conducted a 2-hour programming contest for undergraduate Computer Science students from multiple institutions, where each student was an active member of their institution's computing club. This contest was part of a two-day program organized by \emph{Anonymized} and in each event, students were free to participate individually or in self-formed teams. 27~teams participated in our contest, with each team consisting of students from the same institution. Before the start of the contest, teams were informed that we would request consent to use their anonymized contest data for research purposes at the end of the contest.

\subsection{Contest Format}
In the first 20 minutes, we addressed students as a group to discuss the format and rules for our competitive programming contest. To begin with, we clarified that teams were free to use AI code-generation tools (e.g., ChatGPT, Codeium, or GitHub Copilot) on some or all problems as they wished. Next, we introduced them to the idea of {\pr} Problems. To avoid introducing unfamiliar terms such as `oracle' and `{\pr} Problems', we instead referred to these as `the client' and `Ask the Client Questions' respectively.

As a demonstration, we opened the oracle link for the Practice Problem (Figure~\ref{fig:prac}). To ensure that all contestants were clear about the usual meaning of the term `palindrome' and to familiarize them with the oracle, we demonstrated the usage of the its interface. Each oracle is initialized with an innocuous input that does not reveal any omitted details. For this demonstration, the initial input was the palindrome \verb|'racecar'|. We demonstrated the oracle on two additional innocuous inputs: \verb|'abc'| (return value: \texttt{False}) and \verb|'abba'| (return value: \texttt{True}). 

Next, we showed contestants how to create an initial solution for this problem using an AI code-generation model, Codeium. This model is easy to access via a web-interface\footnote{\url{https://codeium.com/playground}} without requiring any registration. Codeium unsurprisingly suggested Python code that returned the Boolean result of comparing the input string \texttt{s} directly against its reverse \texttt{s[::-1]}.

We then asked contestants to suggest inputs on which we could seek clarifications from the oracle. There were two immediate suggestions made by different students: the singleton letter \verb|'a'| and the empty string. Before querying the oracle on these inputs, we conducted a voice vote on what they expected the answers to be on these inputs based on (a)~the specification, and (b)~the code generated by Codeium. The overwhelming response was \texttt{True} in both cases, and we demonstrated that the oracle agrees with this verdict. After a few seconds, another contestant suggested \verb|'Racecar'| to ``check for case sensitive inputs''. We demonstrated that the oracle returns the somewhat unexpected result \texttt{True} on this input. We further noted that it was ``reasonable for the client to want this behavior, but to forget to mention it'' to the developer. We then asked them to think ``like a developer and seek more clarity''. One student asked whether ``spaces matter'', and we asked them to rephrase their question as a specific input for which we could ask the client for clarification. They suggested \verb|'race car'|, and we showed that the oracle returned \texttt{True} in this case as well. Once again, we highlighted how this was a ``reasonable but unstated'' desire for the client to have.

We then asked a question of our own: ``How do we know that we are done?'' To answer this, we pointed out that contestants would be given a link to a \emph{verifier} for each problem after submitting their first attempts. We demonstrated the verifier for the Practice Problem, whose link is given in the caption of Figure~\ref{fig:prac}. (Similar links appear in the captions of the contest Problems.) We pointed out that each verifier can be used as often as needed, and it will run their code against a set of private test cases (on the basis of which the contest would be judged). However, we clarified that these verifiers merely indicate the \emph{number} of passing test cases, but not the input(s) on which the code failed.

\subsection{Contest Rules}
In the last 10~minutes, we gave contestants specific instructions: they were to work in their current teams on 5~``Ask the Client'' questions for 90~minutes, and we reiterated they were allowed to use any code-generation tools (including Codeium, ChatGPT, and GitHub Copilot). We provided them with a link to a Google Form for their first attempts, which contained links to the oracles for each problem. Thereafter, they would be pointed to a second Google Form containing links to both the oracles and the verfiers, and were free to submit second attempts for any of their failing solutions. After submitting the second form, they would be asked whether \emph{all} members granted consent for us to use their anonymized data for research purposes, or whether \emph{at least one} member did not grant consent. We clarified that their answer would have no bearing on the contest. Except for one team with 2~students, all teams granted consent. Thus, our study is based on 67~students in 26~teams (average: 2.6~students per team).

\section{Results}
\label{sec:methodology}
There were 17~teams with 2~students, 5~teams with 3~students, 2~teams with 4~students, and 1~team each with 1, 5, and 6~students.

\begin{figure*}
 \centering \includegraphics[width=0.6\linewidth]{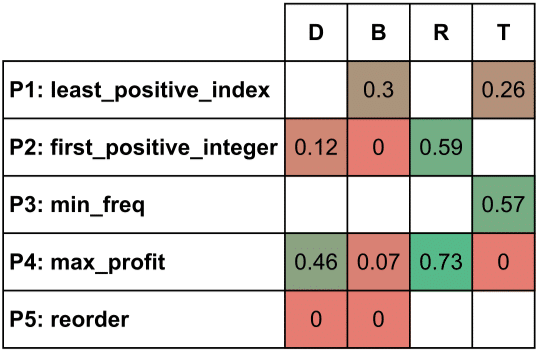}
 \caption{Heatmap of the performance of submissions for each problem against missing details. `D', `B', `R', and `T' respectively represent the four types of omissions: \textit{Definitions},  \textit{Behavior on certain inputs}, \textit{Return type(s)}, and \textit{Tie-break mechanism(s)}. Each cell $(P, X)$ represents the fraction of all submissions for Problem~$P$ that correctly handled omissions of type $X$; this cell is blank when Problem~$P$ does not have an omission of type $X$.
 }
\label{fig: perfheatmap}
\end{figure*}

% \begin{figure*}
%  \centering \includegraphics[width=0.6\linewidth]{images/polar.png}
%  \caption{
%  \todo{Should we put this? I will fix the font sizes}
%  }
% \label{fig: perfheatmap}
% \end{figure*}

\subsection{RQ1: Success at Determining Missing Details}
Overall, we observed that the contestants could determine only about 20\% of the missing details. On average, submissions could only handle 3.88~missing details out of 19 across the five problems. Instead of creating test cases, we evaluated the submitted code using Hypothesis~\footnote{\url{https://hypothesis.readthedocs.io}}, a popular Python library for property-based testing. We created targeted equivalence tests to evaluate each submission against each omission (one at a time). For each test, we carefully specified the class of inputs to test the submission against, ensuring that each test was independent. To ensure that our tests explore the full space of inputs, we ran a ``catch-all'' unrestricted test on submissions after running our targeted tests. However, we found that our targeted tests were collectively exhaustive since no submission passed on all targeted tests but failed on the unrestricted test.

Our evaluation scripts, together with the anonymized contestant submissions as well as the analysis, are available at \url{https://github.com/mrigankpawagi/ProbeableProblems}. 

For Problem 1, around 30\% of the submissions correctly handled the behavior on inputs with more than one instance of the smallest positive integer. Among the rest, two submissions could at least handle a simple case (with the input \texttt{[1, 1]}) although on closer inspection it turned out that this was by coincidence and that these submissions were otherwise logically incorrect. On the other hand, while only 26\% of the submissions correctly handled the behavior on inputs with no positive integers, 63\% of the submissions could at least handle a simple case (with the input \texttt{[]}). This was possibly because the output in the simplest case (which is \texttt{-1}) is conventionally the output on the entire class of such inputs and thus not only could contestants incorrectly assume the intended behavior, but also AI tools are very likely to suggest such code. There was one correct submission to this problem.

For Problem 2, in order to determine whether students figured out the omitted definition of the `first positive integer', we checked the contestant code in three steps. We first checked how they handled single words that contained at most a single positive digit (as its suffix). We then checked how they handled single words that contained several valid prefixes. Finally, we checked how they handled strings containing multiple words. With this characterization, we split the original omission into three sub-omissions -- one corresponding to each of these checks. None of the submissions correctly handled the first, around 66\% correctly handled the second, and 45\% correctly handled the third. Around 59\% of the submissions returned the correct type of values, i.e., integers. Although 59\% of the submissions correctly handled the simplest input (the empty string) without any positive integer, none of the submissions could generalize the notion to the entire class of such inputs.

For Problem~3, 54\% of the submissions correctly handled inputs with multiple least frequent integers -- both in the simplest case (which is \texttt{[1, 0]}) as well as in the general case. It was interesting to note that 93\% of the submissions correctly handled a similar simple input \texttt{[0, 1]} since many contestants either incorrectly assumed that the intended behavior was to pick the first such number in the list, or used an AI tool that produced code with this conventional assumption.

For Problem 4 as well, we split the omission of definition into two separate omissions for the purpose of evaluating contestant code -- the first being that the profit must be non-negative, and the second being the condition \texttt{sell > buy}. We found that while around 87\% of the submissions correctly determined the first property, all the submissions correctly determined the second. Only 37\%, 10\%, and 33\% of the submissions respectively could handle inputs with no non-negative profit, at most zero profit, and with negative list items. Only 30\% of the submissions correctly returned \texttt{None} for the empty list.

For Problem 5, while none of the contestants correctly handled any of the omissions, we found that 12\% of the submissions could handle a simple case with \texttt{NaN} key (with the list \texttt{[0, 1]}), 4\% could handle a simple case with \texttt{NaN} in the list (with the list \texttt{[2, nan, 0]} and the key \texttt{1}), and 28\% could handle a simple case of key values appearing in the list (with the list \texttt{[2, 1, 0]} and the key \texttt{1}).

\subsection{RQ2: Leveraging Verifiers}
 We received a total of 12 second submissions: 2 submissions for Problem~1, 4 for Problem~2, none for Problem~3, 5 for Problem~4, and 1 for Problem~5. In problem 1, one team determined the correct way to handle lists with no positive integers in their second submission. This team correctly handled the simple case in their first submission, and thus feedback from the verifier helped them successfully apply inductive reasoning. In problem 4, one team was able to leverage the verifier to correctly handle inputs with zero profit. On closer observation by taking the \texttt{diff} between their two submissions, we found that they corrected their code by identifying an incorrect \texttt{<} in their code which should have been \texttt{<=}, thereby resolving a tie-break correctly. Three other second submissions for Problem  4 were functionally equivalent to the respective first submissions. All other second submissions handled at least one simple case or class of inputs incorrectly which the respective first submission had handled incorrectly, and did not correctly handle any new input or input class.

\section{Limitations}
Before discussing our findings, we acknowledge two important limitations of our work. First, our categorization of the details omitted during the design of {\pr} Problems (Section~\ref{sec:design_pr}) is \emph{ad hoc}. While these categories seem intuitive to us, we believe that a firmer basis grounded in cognitive and/or linguistic principles should be used (or developed).

Second, while our analysis has focused on the \emph{code} generated by contestants, their interactions with the oracles and verifiers have not been captured. For this study, we hosted these elements on CodeCheck because the constraints of the host event (of which our contest was a part) left us inadequate time to create logins for each team on a server where we could collect data. (The CodeCheck server does not log usage data by design.) We are planning a future study where we will gather and analyze this rich interaction data.

%\section{Discussion}
% While all teams were able to identify some ambiguities in some questions, our analysis of their clarifying questions and their code suggests that they were often unable to recognize several Ambiguous Input Classes (AICs) via their clarifying questions or their code. More worryingly, even teams who succeeded in asking clarifying questions associated with specific AICs were then unable to generate code that either (1)~handled those specific inputs, or (2)~correctly generalized to handling all inputs from those AICs. Further

% It is clear that more feedback (e.g., specific counter-examples) will be helpful.

\section{Discussion and Future Work}
The students who participated in our contest were active members of their institution's computing club, and all teams worked on the problems for at least 85 out of the 90~minutes allotted. Thus, it is reasonable to assume that a good-faith effort was made by most teams. Given this, the low overall performance of teams (Figure~\ref{fig: perfheatmap}) suggests that the contest problems were too challenging for the time allotted. Going forward, we believe that changes to the design of {\pr} Problems and to the contest format are necessary. Further, we believe there are interesting avenues to explore connections between {\pr} Problems in formal educational contexts as well.

\subsection{Reflections on the Contest and {\pr} Problem Design}
Three changes to the design of the contest are immediately warranted. First, providing contestants with a single Practice Problem with just one type of omission appears inadequate. In future iterations of this contest, we will develop a richer set of Practice Problems both for illustration prior to the event and, following~\cite{rein2022competitive}, to help contestants prepare better for the event. Second, given the poor performance on Problem~5 and the very few submissions overall for the second attempt (where the performance on 7 out of 12 resubmissions was worse), it is clear that contestants faced a severe time-crunch. We will organize future contests with fewer problems or provide contestants with more time. Third, we are exploring ways in which the verifiers can provide a limited amount of detailed feedback (e.g., a small number of counter-examples) when the initial submission fails. We believe that this, too, will boost the number of resubmissions.
With regards to the design of {\pr} Questions, we believe that our decision to eliminate \emph{all} details from the complex definition in Problem~2 and from the complex tie-breaker in Problem~4 was poor.
%Further, our oracle at present only provides answers to specific inputs. Once a contestant has formulated an inductive guess for the expected behavior on a general class of inputs, we would like the oracle to be able to handle this general query. Two options that seem promising are: (1)~using the syntax of property-based testing~\cite{10.1145/357766.351266}, and (2)~using natural language.

\subsection{Developing the ability to ask clarifying questions}\label{sec:advanced_ai}
Real-world programming task specifications may be incomplete or may contain statements that are \emph{ambiguous} i.e., they can be interpreted in multiple ways~\cite{berry2004, kamsties2005}. Developers who fail to recognize these deficiencies in the specification may write code that is defective from the client's perspective. Rectifying such defects has the potential to cause expensive delays and may even lead to further defects~\cite{gause1989exploring}. However, a recent study suggests that no special effort is needed to detect such ambiguities ``beyond the normal stakeholders thoroughly discussing the [specification]''~\cite{ribeiro2020}. We believe that CS students must develop the ability to thoroughly discuss specifications, particularly by identifying potential ambiguities and asking clarifying questions. We view {\pr} Problems as one way to develop this ability.

Oh the other hand, we are concerned that AI tools may be able to surpass humans in this ability.  While the performance of LLMs for code-generation have improved significantly, state-of-the-art models are not yet able to match skillful developers~\cite{NEURIPS2023_1b44b878} or participants in competitive programming contests~\cite{jain2024livecodebench}. To improve the performance of these models, a promising line of research is to generate code in careful phases. For instance, ChatDev~\cite{qian2023communicative} mimics the classical 4-stage waterfall model. When presented with a specification that is vague and high-level, ChatDev adopts an LLM-based virtual role and asks clarifying questions to generate a more detailed specification. Similar ideas are used by MetaGPT~\cite{hong2023metagpt}, Alphacodeium~\cite{ridnik2024code}, and reportedly by Devin\footnote{Scott Wu. Introducing Devin, the first AI software engineer.
\url{https://www.cognition-labs.com/introducing-devin}}. A simpler approach by Pawagi and Kumar can solve some {\pr} Problems by leveraging existing AI code-generation tools such as GitHub Copilot~\cite{pawagi2023guardrails}. Specifically, their tool repeatedly invokes the underlying LLM and exploits the fact that missing details are often filled in differently in different solutions. Thus, by identifying inputs on which these solutions differ, it automatically suggests inputs for which clarifications can be sought from a {\pr} Problem oracle.

%\subsection{Connections between {\pr} and Prior Research}
%As noted in  Section~\ref{sec:testing}, we see interesting connections between the skills needed to solve {\pr} Problems and prior research in helping students write good tests. We believe that this is a fruitful area of research.

%Further, we believe that there are interesting questions to explore on how well students comprehend the code~\cite{iticse19wgr}. in a {\pr} Problem when they are writing it manually versus using an AI code-generation tool. For example, if the contestants in each vignette in Section~\ref{sec:pr_intro} written their code manually, would they be more likely to switch between code-level and task-level thinking~\cite{castro2020sigcse}?

\section{Conclusions}
This paper has introduced a novel type of code-writing problem called a {\pr} Problem. The primary motivation for our work is to develop a viable substitute for beginner-level competitive programming problems when the use of AI code-generation tools is permitted. We have showed that carefully constructed {\pr} Problems cannot be solved by modern, easy-to-access AI code-generation tools. Further, we believe that {\pr} Problems can help students develop an important ability: to ask clarifying questions when presented with incomplete or ambiguous task specifications.
%
%There are several variants of {\pr} Problems that may be worth exploring. For instance, the oracles we have considered in this study are ``unconstrained'' i.e., they immediately reveal the expected output for \emph{any} input provided by the contestant. This could adversely impact the ability of contestants to develop their understanding of even the non-ambiguous aspects of the given task specification. Instead, the oracle could be constrained to provide immediate answers only when the input is associated with an omitted detail.
%
Finally, we believe that {\pr} Problems may be interesting for more complex code writing tasks (beyond CS1), and can explore richer ways of introducing ambiguities (e.g., via subtly contradictory statements).

%Bibliography
\bibliographystyle{unsrt}  
\bibliography{sample-base}

\end{document}